# User-driven Intelligent Interface on the Basis of Multimodal Augmented Reality and Brain-Computer Interaction for People with Functional Disabilities

S.Stirenko*, Yu.Gordienko, T.Shemsedinov, O.Alienin,
Yu.Kochura, N.Gordienko

National Technical University of Ukraine
"Igor Sikorsky Kyiv Polytechnic Institute" (NTUU KPI)
Kyiv, Ukraine
*sergii.stirenko@gmail.com

A.Rojbi

CHArt Laboratory (Human and Artificial Cognitions)
University of Paris 8
Paris, France

J.R.López Benito, E.Artetxe González

CreativiTIC Innova SL
Logroño, Spain

*Abstract*—The analysis of the current integration attempts of some modes and use cases of user-machine interaction is presented. The new concept of the user-driven intelligent interface is proposed on the basis of multimodal augmented reality and brain-computer interaction for various applications: in disabilities studies, education, home care, health care, etc. The several use cases of multimodal augmentation are presented. The perspectives of the better human comprehension by the immediate feedback through neurophysical channels by means of brain-computer interaction are outlined. It is shown that brain–computer interface (BCI) technology provides new strategies to overcome limits of the currently available user interfaces, especially for people with functional disabilities. The results of the previous studies of the low end consumer and open-source BCI-devices allow us to conclude that combination of machine learning (ML), multimodal interactions (visual, sound, tactile) with BCI will profit from the immediate feedback from the actual neurophysical reactions classified by ML methods. In general, BCI in combination with other modes of AR interaction can deliver much more information than these types of interaction themselves. Even in the current state the combined AR-BCI interfaces could provide the highly adaptable and personal services, especially for people with functional disabilities.

## I. Introduction

Current investigations of user interface design have improved the usability and accessibility aspects of software and hardware to the benefits of people. But, despite the significant progress in this field, there is still a big work ahead to satisfy requirements of people with various functional disabilities due to lack of adequately accessible and usable systems. It is especially important for persons with neurological and cognitive disabilities. That is why more effective solutions are needed to improve communication and provide the more natural human-to-machine (H2M) and machine-to-human (M2H) interactions, including interactions with their environment (home, office, public places, etc.). The most promising current aims are related to development of technologies aiming at enhancing cognitive accessibility, which allows to improve comprehension, attention, functional abilities, knowledge acquisition, communication, perception and reasoning. One way for achieving such aims is the use of information and communication technologies (ICTs) for development of the better user interface designs, which can support and assist such people in their real environment. Despite the current progress of ICTs, the main problem is the vast majority of people, especially older people with some disabilities, wish to interact with machines in non-obtrusive way and in the most usual and familiar way as much as possible. Meeting their needs can be a major challenge and integration of the newest user interface designs on the basis of the novel ICTs in a citizen-centered perspective remains difficult.

This work is dedicated to the analysis of our previous attempts of integration of some modes of user-machine interaction and the concept of the user-driven intelligent interface on the basis of multimodal augmented reality and brain-computer interaction for various applications, especially for people with functional disabilities. Section II gives the short description of the state of the art in the advanced user-interfaces. Section III contains the concept of the proposed user-driven intelligent interface based on the integration of new ICT approaches and examples of multimodal augmentation developed by authors of this paper. Section IV outlines the opportunities to get neurophysical feedback by brain-computer interaction implemented in several noninvasive BCI-devices presented in the consumer electronics sector of economy. Section V describes the previous results and possibilities to get the better human feedback by integration of multimodal augmentation and brain-computer interaction for the use cases from Section III.

## II. BACKGROUND

The current development of various ICTs, especially related with augmented reality (AR) [1], multimodal user interfaces (MUI) [2], brain-computer interfaces (BCI) [3], machine learning techniques for interpretation of complex signals [4], wearable electronics (like smart glasses, watches, bracelets, heart beat monitors, and others gadgets) [5] open the wide perspectives for development of the user-driven intelligent interfaces (UDII). Some of the most promising UDII are based on psychophysiological data, like heart beat monitoring (HRM), electrocardiogram (ECG), electroencephalography (EEG) and that can be used to infer users' mental states in different scenarios, although they have become more popular recently to evaluate user experience in various applications [6-7]. Moreover, it is possible to monitor and estimate the emotional responses based on these and other physiological measures. For example, galvanic skin response (GSR) gauges the level of emotional excitement or arousal of an individual, which is generally measured by two electrodes on the hands of a participant by the skin conductance level and/or the skin conductance response. These electrodes measure the electrical current differentials stemming from the increase of sweat activity, which often are consequences of the personal excitement [8-9]. Currently, the better user interface designs can be obtained by the development of intelligent, affordable and personalized approaches. They are especially necessary for people with cognitive disabilities to allow them to perform their everyday tasks. Additional needs are related to improvement of their communication channels and uptake of the available and new digital solution and services. The new user interface designs should recognize abilities of customers, detect their behaviors, recognize behavior patterns, and provide feedback in real life environments.

## III. MULTIMODAL AUGMENTATION

The proposed user-driven intelligent interface is assumed to be based on the integration of new ICT approaches and the available ones in favor of the people with functional disabilities. They include the multimodal augmented reality (MAR), microelectromechanical systems (MEMS), and brain-computer interaction (BCI) on the basis of machine learning (ML) providing a full symbiosis by using integration efficiency inherent in synergistic use of applied technologies. The matter is that due to recent "silent revolution" in many well-known ICTs, like AR, ML, MEMS, IoT, BCI, the synergy potential of them becomes very promising. Until recent years AR and BCI devices were prohibitively expensive, heavy, awkward, and especially obtrusive for everyday usage by wide range of ordinary users. The data provided by them were hard to collect, interpret, and present, because of absence of solid and feasible ML methods. But during the last years numerous non-obtrusive AR, BCI, IoT devices become more available for general public and appeared in the consumer electronics sector of economy. At the same time development of MEMS and ML boosted the growth of IoT and wearable electronics solutions proposed on the worldwide scale. Despite these advancements the more effective achievements can be obtained by the proper integration of these ICTs. Below several attempts of such integration of these ICTs are presented, which were laid in the basis of the integral approach.

### A. Tactile and ML for People with Visual Disabilities

Graphical information is inaccessible for people with visual impairment or people with special needs. Studies have demonstrated that tactile graphics is the best modality for comprehension of graphical images for blind users. Usually, graphical images are converted to tactile form by tactile graphic specialists (TGS) involving non-trivial manual steps. Although some exist that contribute to help TGS in converting graphical images into a tactile format, the involved procedures are typically time-consuming, expensive and labor-intensive. In continuation of these efforts the new software program was developed by authors from University of Paris 8 that converts a geographic map given in a formatted image file to a tactile form suitable for people with special needs. The advanced image processing and machine learning techniques were used in it to produce the tactile map and recognize text within the image. The software is designed to semi-automate the translation from visual maps to tactile versions, and to help TGS to be faster and more efficient in producing the tactile geographic map [10-14]. But the further and more effective progress can be achieved when other available ICTs will be integrated. For example, the online feedback for the better comprehension of information can be provided by neurophysical channels by means of BCI and/or GSR interactions. Below in Section IV.A some propositions are given to extend this work for more effective conversion of graphical content to a tactile form.

### B. Visual and Tactile AR for Educational Purposes

As it is well-known, AR consists of the combination of the real world with virtual elements through a camera in real time. This emerging technology has already been applied in many industries. Recently, the effective use of AR in education was demonstrated in order to improve the comprehension of abstract concepts such as electronic fields, and enhance the learning process making education more interactive and appealing for students. One of its main innovations consisted in creation of the totally non-obtrusive Augmented Reality Interface (ARI) by authors from CreativiTIC Innova SL that detects different electronic boards and superposes relevant information over their components, serving also as a guide through laboratory exercises [15-16]. Combination of AR with visual + tactile interaction modes allowed to provide tactile metaphors in education to help students in memorizing the learning terms by the sense of touch in addition to the AR tools. ARI is designed to facilitate learning process and in combination with BCI it can provide specific information about concentration and cognitive load on students. The proper usage of ML methods will allow to conclude and give the contextual AR-based advice to students in classrooms. Below in Section IV.B some ways are described to extend this work for more effective conversion of graphical content to a tactile form with inclusion of other information channels.

### C. TV-based Visual and Sound AR for Home and Health Care

Watching TV is a common activity for every-day life, so adding communication and interactive tools to smart digital TV devices is a good solution to integrate new technologies into ordinary life of people with disabilities not changing their comfort behavior. As far as smart interactive digital TV (iDTV) becomes more popular in homes, they can be used as a core in the current and future tele-healthcare systems. SinceTV system on the basis of the iDTV technology was developed by authors from SinceTV company and National Technical University of Ukraine "Igor Sikorsky Kyiv Polytechnic Institute" to provide visual AR information for various applications [17]. The current prototype includes a iDTV telecare middleware with a hierarchical software stack and structural subsystem. SinceTV is an iDTV technology that can be adapted to improve the lives, for example, of elder people and create integrating modern communication technologies into everyday life and comfortable environment for target users. SinceTV provides interactivity close to real-time; latency minification in reaction fixing; synchronization using ACR (Audio Recognition Content); high load scalability up to 10 million concurrent connections (potentially linear grow); the second screen concept. SinceTV allows you to add interactive AR data to video and audio streams, linking two points, not only through the media devices, but also provides facilities for distributed interactive applications. Such a set of features in combination AR-based feedback can be useful for health care purposes, for example, for activity measurement and health state estimation via vision-based algorithms. Below in Section IV.C some potential directions of the further development of this iDTV system are given.

### D. Visual AR + Wearable Electronics for Health Care

The standard cardiology monitoring can show the instant state of cardiovascular system, but unfortunately, cannot estimate the accumulated fatigue and physical exhaustion. Errors due to fatigue can lead to decrease of working efficiency, manufacturing quality, and, especially, workplace and customer safety. Some specialized commercial accelerometers are used to record the number of steps, activities, etc. [18]. However, they are quite limited to assess the health state and measure accumulated fatigue. The new method was proposed recently by authors from National Technical University of Ukraine "Igor Sikorsky Kyiv Polytechnic Institute" to monitor the level of currently accumulated fatigue and estimate it by the several statistical methods [19].

The experimental software application was developed and used to get data from sensors (accelerometer, GPS, gyroscope, magnetometer, and camera), conducted experiments, collected data, calculated parameters of their distributions (mean, standard deviation, skewness, kurtosis), and analyzed them by statistical and machine learning methods (moment analysis, cluster analysis, bootstrapping, periodogram and spectrogram analyses). Various gadgets were was used for collection of accelerometer data and visualization of output data by AR. Several "fatigue metrics" were proposed and verified on several focus groups. The method can be used in practice for ordinary people in everyday situations (to estimate their fatigue, give tips about it and advice on context related information) [20]. In addition to this, the more useful information as to fatigue can be obtained by estimation of the level of user concentration to the external stimuli by brain-computer interaction that is described below.

By EEG measurements they can determine different psychophysiological states, such as attention, relaxation, frustration, or others. For example, MindWave Mobile by Neurosky allows you to determine at least two psychological states: concentration ("attention") and relaxation ("meditation"). The exposure of a user to different external stimulators will change both levels of the psychological states collected by this device. For example, this method can estimate: if the user is calm, then the relaxation ("meditation") will be high and the concentration ("attention") will be low. The consumer BCI-devices have various number of EEG channels, types of EEG connection with human surface, different additional sensors, and their price depends on their possibilities (see Table 1). Fortunately, all of them have additional documentation for software developers and related software development kits (SDKs), which allow external developers to propose their own solutions and make research.

## IV. NEUROPHYSICAL FEEDBACK

Brain-Computer Interfaces (BCIs) are widely used to research interactions between brain activity and environment. These researches are often centered on mapping, assisting, augmenting, or repairing human cognitive or sensory-motor functions. The most popular signal acquisition technology in BCI is based on measurements of EEG activity. It is characterized by different wave patterns in the frequency domains or "EEG rhythms": alpha (8 Hz - 13 Hz), SMR (13 Hz - 15 Hz), beta (16 Hz - 31 Hz), Theta (4 Hz - 7 Hz), and Gamma (25 Hz - 100 Hz). They are related with various sensorimotor and/or cognitive states, and translating cognitive states or motor intentions from different rhythms is a complex process, because it is hard to associate directly these frequency ranges to some brain functions. Some consumer EEG solutions, such as the MindWave Mobile by Neurosky, Muse by InteraXon, Emotiv EPOC by Emotiv and the open-source solutions like OpenBCI become available recently and can be used to assess emotional reactions, etc (see Figure 1).

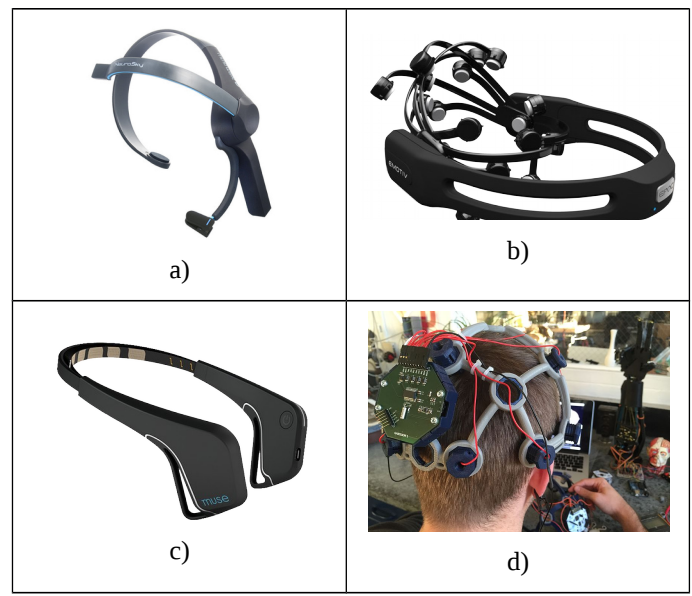

Fig. 1. The noninvasive BCI-devices presented in the consumer electronics sector of economy: a) Mind Wave Mobile by NeuroSky (http://neurosky.com); b) EPOC by Emotiv (https://www.emotiv.com); c) Muse by InteraXon (http://www.choosemuse.com); d) Ultracortex Mark III by OpenBCI (http://openbci.com).

TABLE I. COMPARISON OF SOME CONSUMER BCI DEVICES

| Device (Company) | EEG channels | EEG connection | Additional sensors | Price, $ |
|---|---|---|---|---|
| MindWave Mobile (NeuroSky) | 1 | dry | accelerometer | 89 |
| EPOC/Insight (Emotiv) | 5/14 | wet/dry | accelerometer, gyroscope, magnetometer | 300-800 |
| Muse (InteraXon) | 5 | dry | accelerometer | 249 |
| OpenBCI (open source) | 4/8/12 | dry/wet | EMG/ECG | 750-1800 |

Unfortunately, all consumer BCI-devices are specialized setups without sound, visual, tactile and other feedbacks in reaction to the external stimuli. Their wet contact should be used with the specialized gel, but the more feasible practical applications are possible mainly on the basis of the dry contacts. Additional obstacle is that the consumer BCI-devices and the related software are proprietary solutions, which cannot be easily developed, adopted, and used by the third-parties, especially in education and research purposes. In this connection availability of Open Brain Computer Interface (OpenBCI) open the ways for science advancements by openly shared knowledge among the wider range of people with various backgrounds. It will allow them to leverage the power of the open source paradigm to accelerate innovations of H2M and M2H technologies. For example, the available OpenBCI hardware solutions like Ganglion (with the 4-channel board) is suitable for low-cost research and education, and Cyton (with the 8-16 channel boards) provides the higher spatial resolution and enables more serious research.

## V. USE CASES FOR AR-BCI INTEGRATION

Before in Section III several examples of effective usage of the various ICTs combinations and interaction channels were demonstrated on the application level. The similar approach was presented recently as the augmented coaching ecosystem for non-obtrusive adaptive personalized elderly care on the basis of the integration of new and available ICT approaches [21]. They included multimodal user interface (MMUI), AR, ML, IoT, and machine-to-machine (M2M) interactions based on the Cloud-Fog-Dew computing paradigm services.

Despite the current progress in the above mentioned attempts to combine the available AR modes, the most promising synergy can be obtained by online EEG and GSR data monitoring, processing, and returning as AR feedback. The general scheme of multimodal interactions and data flows is shown in Figure 2: collection of EEG reaction from user by various available consumer BCI-devices (Figure 1, Table 1), ML data processing, and return output data to user as AR feedback by available AR-ready gadgets. The crucial aspect is to avoid the obtrusive way of usage of the current BCI-gadgets, which can be much more inappropriate by users, if they will be equipped by additional AR-features. But the current progress of microcontrollers, sensors, and actuators allow to use the combination of low cost contacts, microcontrollers with low energy Bluetooth or Wi-Fi wireless networking, ear phones for sound AR and LEDs for visual AR on the ordinary glasses instead of "Terminator"-like bulky and awkward specialized devices (Figure 2).

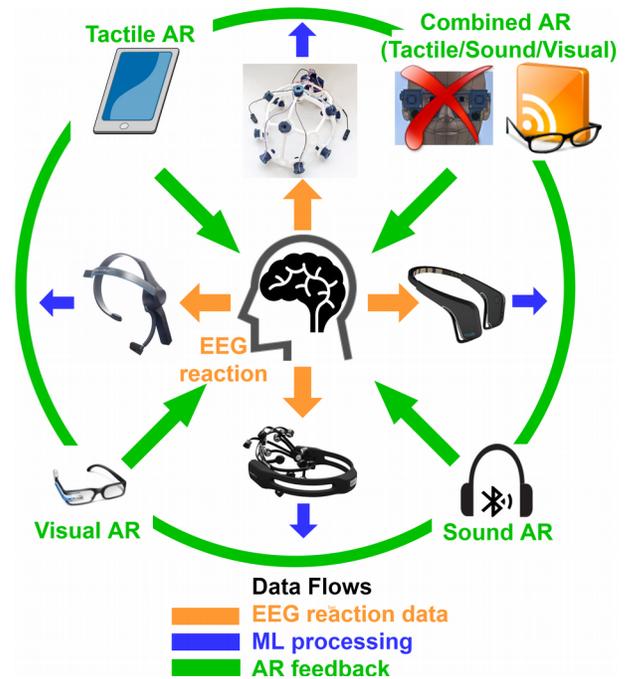

Fig. 2. The general scheme of multimodal interactions and data flows: collection of EEG reaction from user by various available BCI-ready devices (see Table 1), ML data processing, and return output data to user as AR feedback by available AR-ready gadgets.

This general concept of multimodal integration was verified by the experimental setup (Figure 3). It includes smart glasses Moverio BT-200 by EPSON as a visual AR interaction channel for the controlled cognitive load (set of mathematical exercises) and a collector of accelerometer data; neurointerface MindWave by NeuroSky as a BCI-channel and collector of EEG-data; and heart monitor UA39 by Under Armour as a collector of heartbeat data. The setup can collect time series of several parameters: the subtle head accelerations (like tremor characterizing stress), EEG-activity, and intervals of heartbeats on the scale of milliseconds. The statistical methods were used to find correlations between these time series for various conditions, and machine learning methods were used to determine and classify various regimes.

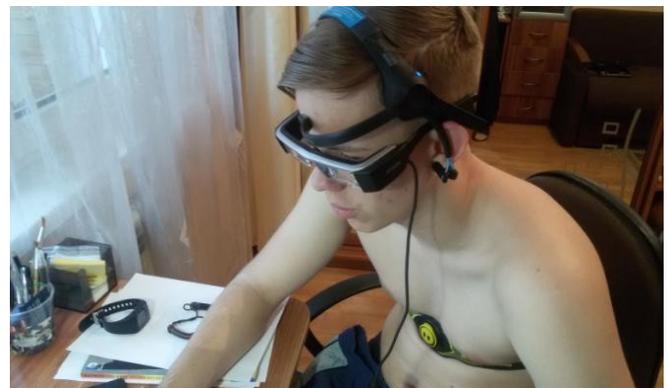

Fig. 3. Smart glasses Moverio BT-200 by EPSON (on the eyes) as an AR interaction channel and a collector of accelerometer data; neurointerface MindWave by NeuroSky (on the head with the blue band) as a collector of EEG-data; and heart monitor UA39 by Under Armour (on the breast with the yellow spot) as a collector of heartbeat data.

The previous experiments with the general concept and setup allowed us to propose several possible applications for the integration of AR channels with BCI technologies to provide the direct neurophysical feedback to users, which are discussed below. The general idea of the AR-BCI integration is based on the establishments of multimodal interactions and data flows, where all EEG reactions from the user observed by various available BCI-ready devices (Figure 1, Table 1) or combined AR-BCI devices are gathered and then they are processed by ML methods on the supportive devices (smartphone, tablet, etc) in non-obtrusive way. The essence of the AR-BCI integration consists in the real-time return of the obtained output data of neurophysical nature to user as AR feedback by available AR-ready gadgets (through sound, visual, and tactile AR channels).

### A. BCI for People with Visual Disabilities

The application mentioned in Section III.A was developed on the basis of the advanced image processing and ML techniques to produce the tactile map and recognize text within the image. The available tools allow to automate the conversion of a visual geographic map into tactile form, and to help tactile graphics specialists be more efficient in their work. The development of these tools has benefited from feedback from specialists of National Institute for Training and Research for the Education of Disabled Youth and Adapted Teaching (INS HEA) [22] and from volunteers. The previous analysis of the available mode of operation and possible improvements opened the following ways for improvement of the cognitive abilities during reading of various multimedia materials by tactile contacts. Even the low end consumer and open-source BCI-devices (like 1-channel MindWave Mobile by Neurosky or 4-channel Ganglion by OpenBCI) can differentiate, at least, two (or four) psychological states: concentration and relaxation. The exposure of a user to different external stimulators (for example, through various tactile interactions) will change both levels of the psychological states collected by this device. For example, this method can estimate: if the user has a good tactile contact, then the concentration will be high, and the relaxation will be low. In such a case the measure of the proper tactile contact through such BCI feedback can be helpful for online estimation of the automatic conversion of a visual geographic map into tactile form. The future development of this solution should recognize user's abilities and be able to detect behaviors and recognize patterns, emotions and intentions in real life environments. In this case a mix of technologies such ML and tactile interaction with BCI will profit from the immediate feedback on the basis of the actual neurophysical reactions classified by ML methods.

### B. BCI for Educational Purposes

The several successful attempt to combine AR and other interaction modes were demonstrated in Section III.B on the basis of tactile haptic pen and tactile feedback analysis in education. There the simple classification of functions was used to develop tactile metaphors targeted to help students memorize the learning terms by the sense of touch in addition to the AR tools. The objective of the tactile accessory was to create different vibrotactile metaphors patterns easy to distinguish without ambiguity. Experiments shown that the vibrotactile feedback is well perceived by the user thanks to the wide range of frequency and amplitude vibration provided by the innovative combination of vibrotactile actuators. The next important step can be to measure if metaphors are relevant and effectively help students to memorize learning concepts (especially in lifelong learning) by additional neurophysical feedback by BCI along with tactile interaction. Besides supplying the localization of the zone of interest for the AR process, the role of the BCI is to provide "the positive feedback for satisfactorily recognized metaphors". The real objective of this AR-BCI system is to allow the user to see his/her own the physical feelings about the zone of interest. This approach, based on a "tangible" and "thinkable" object, has to incite the user to explore an invisible notions and ambience which compose the zone of interest.

### C. BCI for TV-based Home Care

SinceTV system on the basis of the iDTV technology (described in Section III.C) provides interchange of generalized data structures of the various types like interactive questions and answers (with various input devices) and values obtained from the sensors, electronic equipment and different devices. It can process calls, events and data synchronization for distributed applications. In combination with neurophysical data obtained from users by BCI-devices in intrinsically interactive mode of operation, this technology can be helpful for much better communication of elderly people with each other, relatives, caregivers, doctors, social workers. Communication includes multi-platform applications: mobile, web, desktop and specialized interactive TV interface with voice and video input for target users. TV-based AR and BCI can provide the quite different way for estimation of concentration level of elderly people during their sessions of watching TV programs and shows, selection of food, goods, and services. This BCI feedback information can be invaluable for remote diagnostics and medical devices control.

### D. BCI for Wearable Health Care

The more sophisticated estimation of various types of everyday and chronic fatigue (including mental, and not only physical) can be obtained by measuring the level of user concentration to the external stimuli by the low end consumer and open-source devices even. The data from sensors like accelerometer were collected, integrated, and analyzed by several statistical and machine learning methods (moment analysis, cluster analysis, principal component analysis, etc.) (Fig. 4). The proposed method consists in monitoring the whole spectrum of human movements, which can be estimated by Tri-Axial Accelerometer (TAA), heartbeat/heartrate (HB/HR) monitor, and BCI (optionally, + by synergy with other sensors in the connected smartphone and data on ambient conditions in the smartphone). The main principle is the paradigm shift: to go from "raw output data" (used in many modern accelerometry based activity monitors) to the "rich post-processed (and, optionally, ambient-tuned) data" obtained after smart post-processing and statistical (moment/cluster/bootstrapping) analyses with much more quantitative parameters.

The hypothesis 1 (physical activity can be classified) and hypothesis 2 (fatigue level can be estimated quantitatively and distinctive patterns can be recognized) were proposed and proved, and due to shortage of space here the details are given elsewhere [23]. Several "fatigue metrics" were proposed and verified on several persons of various age, gender, fitness level, etc. Correlation analysis of the moments (mean, standard deviation, skewness, kurtosis) of statistical distribution of acceleration values allowed us to determine the pronounced correlation between skewness and kurtosis for the states with high level of physical fatigue: after physical load (Fig.5a) and in the very end of the day (Fig.5d).

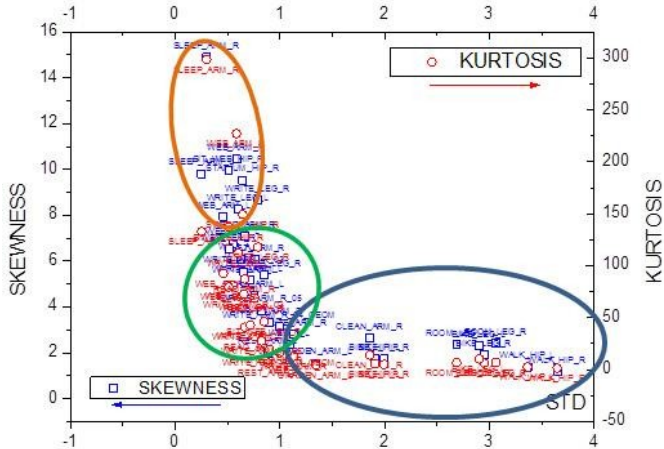

Fig. 4. Multiparametric moment analysis: activities can be classified in more details, i.e. divided into groups (colored ellipses) with the similar values of the acceleration distribution parameters: the active (sports, housework, walking) (blue ellipse), moderate (writing, sitting) (green ellipse) and passive (web surfing, reading, sleeping) (brown ellipse) behavior.

The similar ideology was applied for estimation of the workload during exercises and its influence on heart. The crucial aspects of this approach are as follows:

(1) the absolute values of heart rates (heartbeats) for the same workload (Fig. 6) are volatile and sensitive to the person (age, gender, physical maturity, etc.) and its current state (mood, accumulated fatigue, previous activity, etc.) – what should be done: in contrary, their distributions should be used here instead;

(2) the heart rate values are actually integer values with 2-3 significant digits and not adequately characterize the volatile nature of heart activity (because the heart rate is actually the reverse value of the heartbeat multiplied by 60 seconds and rounded to integer value) – what should be done: in contrary, heartbeats in milliseconds should be used, because they contain 3-4 significant digits and their usage gives 10 times higher precision;

(3) the actual influence of workload on heart, accommodation abilities of heart, and fatigue of heart were not estimated before – what should be done: the results from statistical physics as to the critical phenomena and processes in the context of heart activity should be used.

This method allows us to determine the level of fitness from the moments diagrams of the heartbeat distribution functions vs. exercise time for various workloads: well-trained person (YU, male, 47 years) (Fig.7a) and low trained person (NI, male, 14) (Fig.7b).

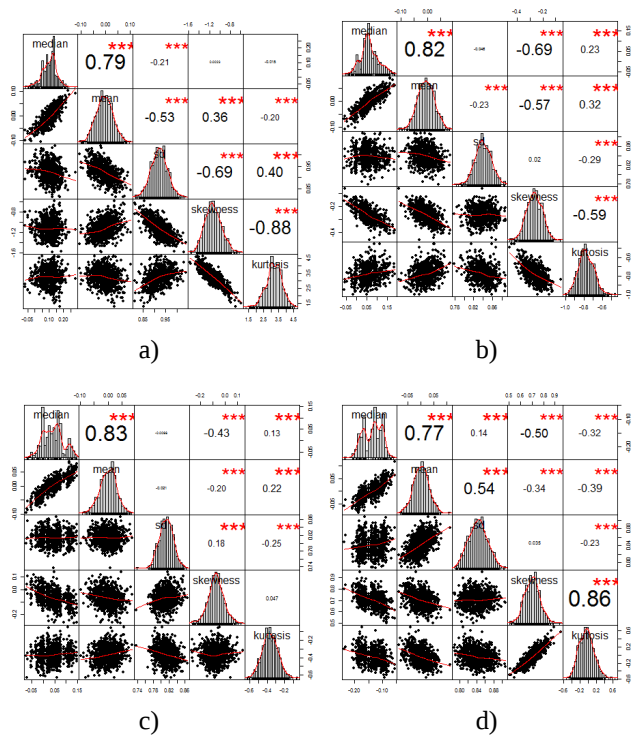

Fig. 5. Correlation analysis of the moments (mean, standard deviation, skewness, kurtosis) for statistical distribution of acceleration values for states with different levels of physical fatigue: (a) wake-up state, (b) after physical load (10 km of skiing), (c) rest state after lunch, (d) in the very end of the day.

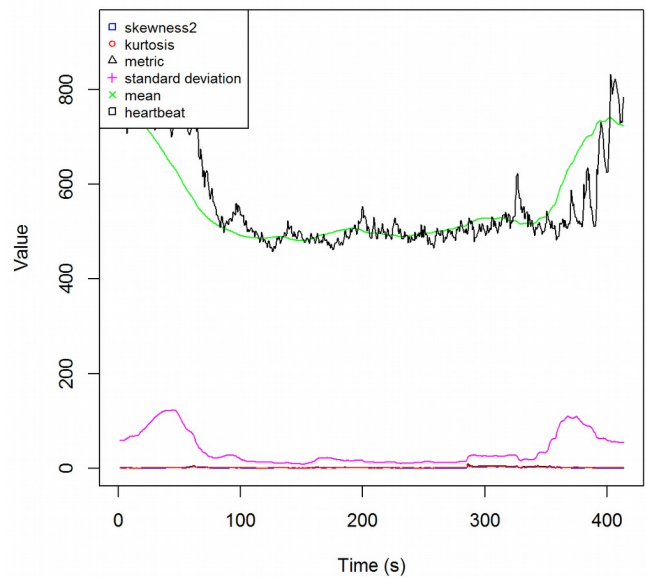

Fig. 6. Statistical parameters used for estimation of heartbeat/heartrate activity during exercises (heartbeat at walking for the well-trained person, male, 47 years). Legend: top black line — heartbeat itself, green — moving mean, magenta — standard deviation, low black line — metric, red — kurtosis, blue — skewness. (Kurtosis and skewness are not seen here, because of their low values. Please, see the next plots below.) The exercise was like: 1 min of rest + 5 min of walking with velocity 6.75 m/s + 1 min of rest.

The exercise squats were performed up to fatigue. Size of blue symbols (the current moments of heartbeat distribution)

increases with time of exercise. The distribution functions for the higher fitness (Fig.7a) have tendency to slower and nearer movement of points with time of exercise (i.e. shift to the higher values of moments), and Distribution functions for the lower fitness (Fig.7b) have tendency to much faster and farther movement.

of exercise. The exercises were performed up to fatigue and denoted as "0.5 kg" — 0.5 kg dumbbell curl for biceps, "1 kg" — 1 kg dumbbell curl for biceps, "3 kg" — 3 kg dumbbell curl for biceps. Again, the distribution functions for the higher fitness have tendency to slower movement of points with time of exercise (i.e. shift to the higher values of moments), and the distribution functions for the lower fitness have tendency to much faster movement. And the Distribution functions for the higher workload (weight of dumbbell, here) have tendency to much faster movement of points with time of exercise (i.e. shift to the higher values of moments).

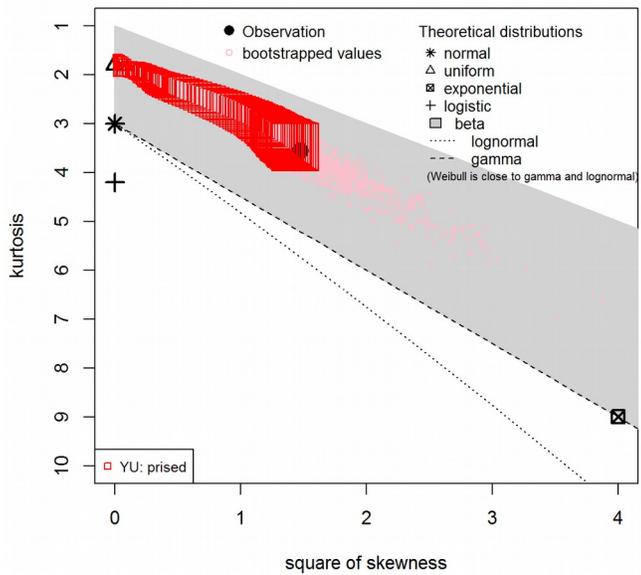

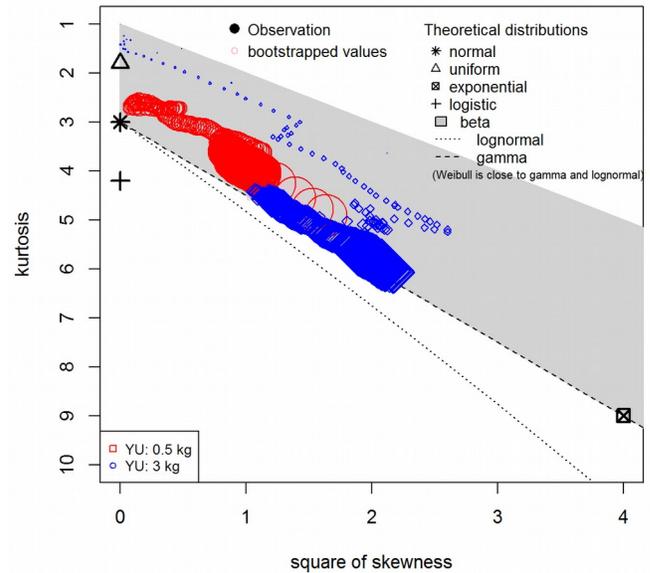

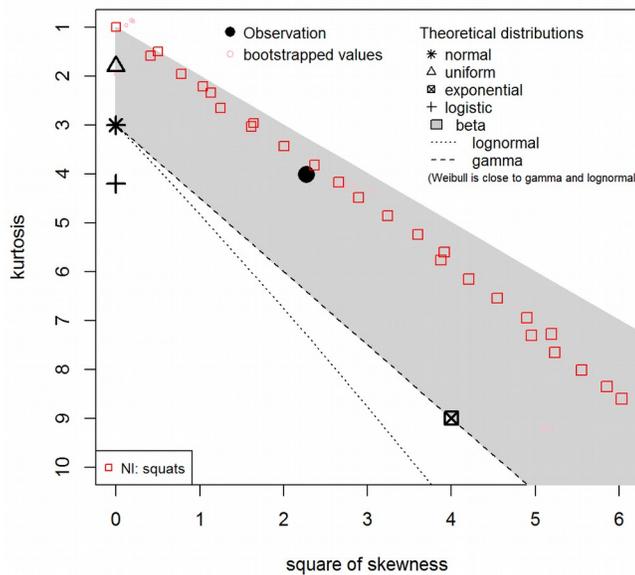

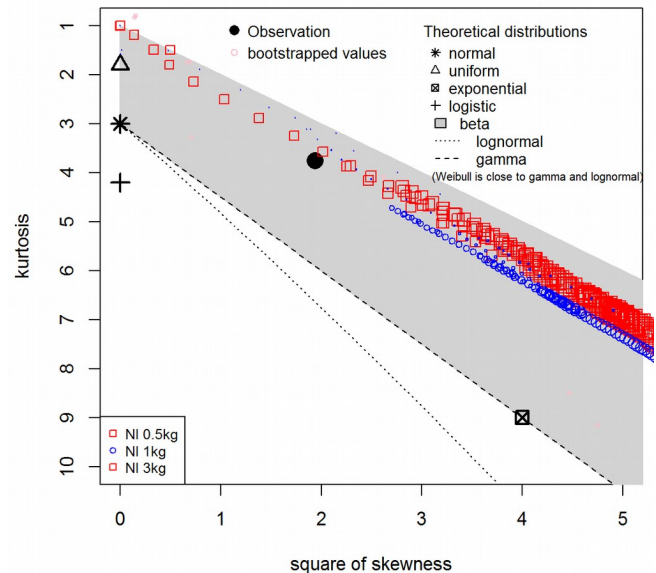

**Fig. 7.** Moments diagrams of the heartbeat distribution functions vs. exercise time for squats of: a) well-trained person (YU, male, 47 years) and b) low trained person (NI, male, 14). The exercise squats were performed up to fatigue. Legend: Size of blue symbols (current moments of heartbeat distribution) increases with time of exercise.

This method allows us to determine the level of fitness for other types of exercise (for example, dumbbell curl for biceps here) from the moments diagrams of the heartbeat distribution functions vs. exercise time for various workloads: a) well-trained person (YU, male, 47 years) and b) low trained person (NI, male, 14) (Fig.8). The size of symbol increases with time

**Fig. 8.** Moments diagrams of the heartbeat distribution functions vs. exercise time for various workloads (see legend): a) well-trained person (YU, male, 47 years) and b) low trained person (NI, male, 14). Legend: size of symbol increases with time of exercise; exercises were performed up to fatigue and denoted as "0.5 kg" — 0.5 kg dumbbell curl for biceps, "1 kg" — 1 kg dumbbell curl for biceps, "3 kg" — 3 kg dumbbell curl for biceps.

From the empirical point of view, the various metrics can be created on this basis, for example, "METRIC" as the distance from the uniform distribution to the current position (Fig.9a), or "METRIC3" as the distance from the normal distribution to the current position (Fig.9b). These metrics allows us to characterize and differentiate the workload levels and recovery phases, for example, from the previous exercise with various walking and jogging velocities. The slopes of metric increase and decrease can be used to characterize the accommodation and recovery levels during these exercises. Note: the initial sharp red peak for the highest possible load corresponds to the recovery phase after previous exercise.

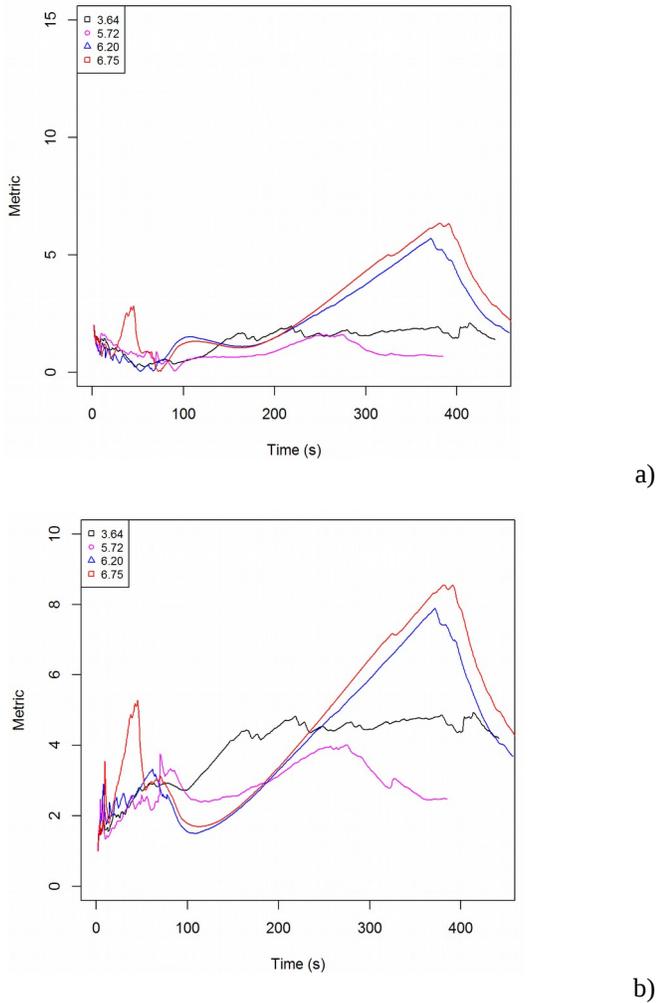

a)

b)

Fig. 9. Metrics of the heartbeat distribution functions vs. walking velocities for the well-trained person (YU, male, 47 years): a) METRIC as the distance from the normal distribution on the moments diagram, b) METRIC3 as the distance from the uniform distribution on the moments diagram. The exercise was like: 1 min of rest + 5 min of walking + 1 min of rest. Legend: black line — 3.64 m/s (very low load), magenta line — 5.20 m/s (comfort load), blue line — 6.20 m/s (high load), red line — 6.75 m/s (highest possible load).

The similar measurements of EEG brain activity by BCI-channel show the same output as to paradigm of usage distributions instead of separate absolute values provided by sensors. The raw absolute values of EEG activities measured as attention (ATT), relaxation (REL), and eye blink levels (EYE)

(Fig.10a) cannot provide the useful information, and it is hard to find any correlation between these levels (Fig.10b).

But similar approach on the basis of distributions and their moments can provide much more valuable information. For example, the similar "METRIC" as the distance from the uniform distribution to the current position, allow to find visual qualitative (Fig.11) and quantitative numerical correlation (for attention and relaxation metrics) and anticorrelation for (for eye blink and relaxation metrics).

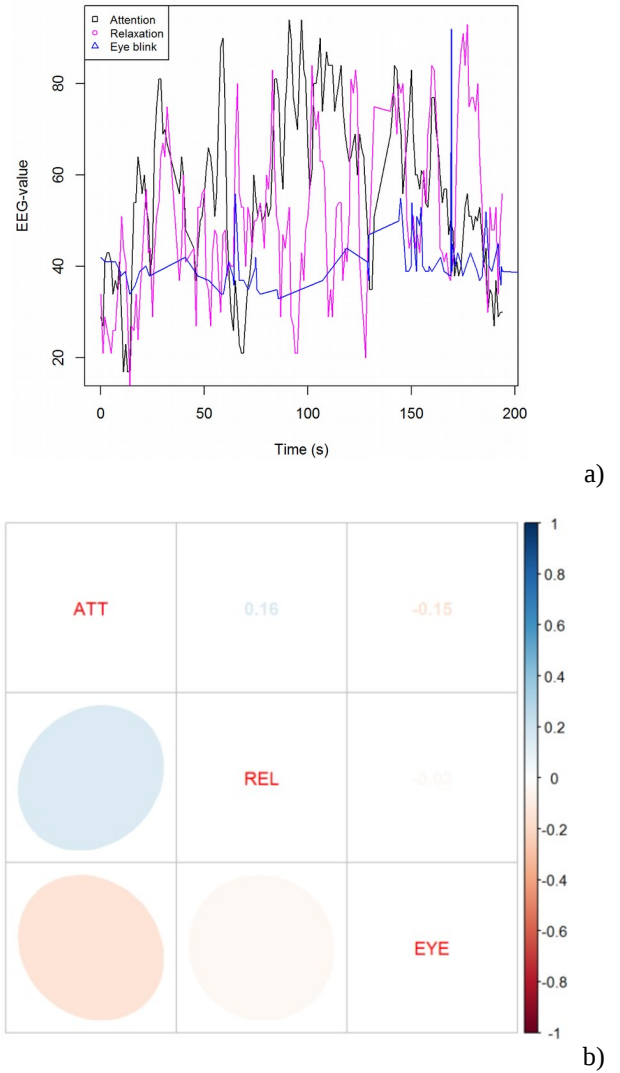

a)

b)

Fig. 10. EEG activities measured as attention (ATT), relaxation (REL), and eye blink levels (EYE): (a) absolute values, (b) their correlation matrix.

At the moment, these findings are isolated from each other, because they were measured separately, but the mush bigger potential could be foreseen if they will be combined [23]. In combination with the proposed fatigue metrics, BCI can provide the actual neurophysical feedback for users in the wearable electronics, like usual glasses (cap, hat, band, etc.) with the attached BCI-contacts, heart rate monitors, other microcontrollers with low energy Bluetooth or Wi-Fi wireless networking, simplified visual AR (like LEDs) and/or sound AR (ear phones) indicators (Figure 3). It should be noted that at the moment this combination of AR+BCI and its hardware concept is on the stage of estimation of its general feasibility on the basis of the currently available ICTs. And it cannot be used for diagnostics of health state (including fatigue) in any sense, because involvement of various expertise (including medicine, psychology, cognition, etc.) is necessary and related specific research should be carried out.

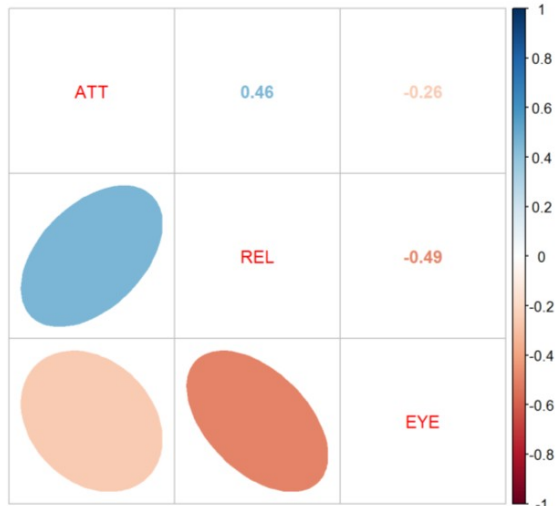

Fig. 11. EEG activities measured as attention (ATT), relaxation (REL), and eye blink levels (EYE): (a) metrics on the basis of the moments of distributions of absolute values, (b) their correlation matrix.

## VI. CONCLUSIONS

In general, brain-computer interaction in combination with other modes of augmented interaction might deliver much more information to users than these types of interaction themselves. BCI provides quantitative measures of relevant psychophysiological actions and reactions and allow us to truly determine what was perceived or felt while our sensorimotor and/or cognitive systems are exposed to a stimulus. In the context of home and health care for people with functional disabilities, these quantitative measures (like GSR, EEG, ATS, etc.) cannot replace the current methods of evaluation and proactive functions, but they can complement and enhance them. But the use of other sensor data along with EEG activity can be very meaningful in the context of evaluating the user mental and physical state. The proposed user-driven intelligent interface on the basis of multimodal augmented reality and brain-computer interaction can be useful for various mentioned applications (education, lifelong learning, home care, health care, etc.). It could improve communication and interaction capability of people with disabilities and facilitate social innovation. It should be noted that in the context of estimating scene geometry and complex relationships among objects for autonomous vehicle driving such an intelligent interface can be useful for providing the instant feedback from humans for creation and development of the better machine learning approaches for visual object recognition, classification, and semantic segmentation [24]. Usage of open-source hardware and software solutions (like OpenBCI) could allow developers to leverage the more affordable technologies and products to support interactions for people with disabilities. It should be noted again, that it cannot be used for diagnostics of health state in any sense, because synergy of various expertise (including relevant disciplines like psychology, cognition, disability, etc.) is necessary for the further development and test of the proper solutions, models and algorithms to improve information extraction from neurophysical signals. But even in the current state this new generation of combined AR-BCI interfaces could provide the highly adaptable and personalisable services to individual contexts, especially for people with functional disabilities.

### ACKNOWLEDGMENT

The work was partially supported by Ukraine-France Collaboration Project (Programme PHC DNIPRO) (http://www.campusfrance.org/fr/dnipro), Twinning Grant by EU IncoNet EaP project (http://www.inco-eap.net/), and EU TEMPUS LeAGUe project (http://tempusleague.eu).